\documentclass[12pt]{article}
\usepackage{amsmath}
\usepackage{graphicx}
\usepackage{enumerate}
\usepackage{natbib}
\usepackage{booktabs, caption, makecell}
\usepackage{subcaption}
\usepackage{multirow}
\usepackage{bm}
\usepackage{authblk}
\usepackage[dvipsnames]{xcolor}

\usepackage{enumitem}
\setlist[itemize]{leftmargin=*}

\usepackage{stackengine}

\newcommand{\blind}{1}

\begin{document}

\def\spacingset#1{\renewcommand{\baselinestretch}%
{#1}\small\normalsize} \spacingset{1}

%%%%%%%%%%%%%%%%%%%%%%%%%%%%%%%%%%%%%%%%%%%%%%%%%%%%%%%%%%%%%%%%%%%%%%%%%%%%%%

\if1\blind
{
  \title{\bf Two-level D- and A-optimal main-effects designs with run sizes one and two more than a multiple of four}
  \author{Mohammed Saif Ismail Hameed$^1$}
    \author{José Núñez Ares$^1$}
    \author{Eric D. Schoen$^{1}$}
    \author{Peter Goos$^{1,2}$}
    \affil{$^1$Department of Biosystems, KU Leuven, Leuven, Belgium}
    \affil{$^2$Department of Engineering Management, University of Antwerp, Antwerp, Belgium}
    %   \author{Author 1\thanks{
    % The authors gratefully acknowledge \textit{please remember to list all relevant funding sources in the unblinded version}}\hspace{.2cm}\\
    % Department of YYY, University of XXX\\
    % and \\
    % Author 2 \\
    % Department of ZZZ, University of WWW}
  \maketitle
} \fi

\if0\blind
{
  \bigskip
  \bigskip
  \bigskip
  \begin{center}
    {\LARGE\bf Two-level D- and A-optimal main-effects designs with run sizes one and two more than a multiple of four}
\end{center}
  \medskip
} \fi

\bigskip
\begin{abstract}
\noindent For run sizes that are a multiple of four, the literature offers many two-level designs that are D- and A-optimal for the main-effects model and minimize the aliasing between main effects and interaction effects and among interaction effects. For run sizes that are not a multiple of four, no conclusive results are known. In this paper, we propose two algorithms that generate all non-isomorphic D- and A-optimal main-effects designs for run sizes that are one and two more than a multiple of four. We enumerate all such designs for run sizes up to 18, report the numbers of designs we obtained, and identify those that minimize the aliasing between main effects and interaction effects and among interaction effects. Finally, we compare the minimally aliased designs we found with benchmark designs from the literature.
\end{abstract}

\noindent%
{\it Keywords:}  design enumeration, minimal aliasing, non-isomorphic design, optimal experimental design
\vfill

\newpage
\spacingset{1.9} % DON'T change the spacing!

\spacingset{1}
\section{Introduction} \label{sec:intro_1mod4}
\spacingset{1.9}

%Body of paper.  Margins in this document are roughly 0.75 inches all around, letter size paper.

A common criterion to select an experimental design with $N$ runs and $k$ factors for {some research project} is the determinant of the information matrix $\mathbf{X'X}$, where $\mathbf{X}$ is the model matrix corresponding to the linear regression model under consideration. A design that achieves the maximum value for this criterion is labeled D-optimal where ‘D’ stands for determinant \citep{goos2011optimal}. D-optimal designs minimize the generalized variance of the estimates of the effects in the regression model. Such designs also minimize the volume of the confidence ellipsoid of the parameters \citep{atkinson2007optimum}. An alternative design selection criterion is the trace of $(\mathbf{X'X})^{-1}$. A-optimal designs minimize this trace and consequently have the lowest average variance for all estimates of the effects in a given model \citep{atkinson2007optimum}.

Given the importance of experimental factors’ main effects and the possible importance of their interactions, it would be ideal if an experimental plan were D- and A-optimal for the complete main-effects-plus-interactions model. However, for $k$ factors, D- and A-optimal experimental plans for this model require at least $1+ k+\binom{k}{2}$ runs, so that they are often too costly to be used in practice. {For this reason, it is common to optimize the experimental design for the main effects, while making sure that the bias in the main-effect estimates is minimized and the information on the interactions is maximized.}

For most run sizes and numbers of factors, D-optimal designs for the main-effects model are also A-optimal for that model, regardless of whether we restrict ourselves to the class of two-level designs or to designs for quantitative factors that can take any value on an interval \citep{jones2021optimal}. This implies that, even in the event the factors can, in principle, take any value on an interval, only two levels are used for each factor in most D- and A-optimal main-effects designs. In this paper, we refer to all two-level designs that are both D- and A-optimal for the main-effects model as DA designs. We denote the two levels of the designs' factors by $-1$ and $1$.

For any given pair of number of runs $N$ and number of factors $k$, many different DA designs exist. Some of these DA designs are isomorphic to each other, because it is possible to obtain one design from another by permuting its columns, permuting its rows, and/or switching the signs of the levels of one or more of its factors. {All the designs that are isomorphic to each other form an isomorphism class. Designs of the same isomorphism class possess the same statistical qualities.} Therefore, it makes sense to study only sets of non-isomorphic DA designs.

The literature on design of experiments contains multiple studies of non-isomorphic DA designs for cases where the run size $N$ is a multiple of four. For that type of run size, all DA designs are orthogonal arrays and the research performed can be considered definitive, in the sense that complete catalogs of non-isomorphic DA designs have been generated and studied, and the DA designs that minimize the aliasing between main effects and interactions as well as among the interactions are known for run sizes up to 24. References include \citet{LIN1993147}, \citet{deng1999generalized} and \citet{schoen2017two}. For run sizes $N$ that are not multiples of 4, hardly any research has been performed concerning non-isomorphic DA designs. One exception is the work of \citet[ZM]{zhang2013minimum}, who sought DA designs that minimize the aliasing between main effects and higher-order interaction effects for run sizes that are one more than a multiple of 4. Another exception is the work of \citet[HM]{huda2013two}, who performed similar research for run sizes that are two more than a multiple of four. To the best of our knowledge, there are no published studies of non-isomorphic DA designs for run sizes that are three more than a multiple of four.

Unlike the work on DA designs for run sizes that are multiples of 4, the work of ZM and HM for run sizes that are one and two more than a multiple of 4 cannot be considered definitive. This is due to the fact that these authors did not consider complete sets of non-isomorphic DA designs. More specifically, ZM only considered DA designs that can be obtained by adding one row to an orthogonal array, and HM only considered DA designs that can be obtained by adding two rows to an orthogonal array. In this paper, we show that, for multiple combinations of the number of runs $N$ and the number of factors $k$, better DA designs exist than those found by ZM and HM, in the sense that they involve less aliasing between the main effects and the interactions and/or among the interactions. To identify the best possible DA designs, we first developed two enumeration algorithms to generate complete sets of non-isomorphic DA designs for run sizes that are one and two more than a multiple of 4. {Next, we quantified the aliasing between the effects of interest in these complete sets of DA designs with the G- and G$_2$-aberration criteria (\citet{deng1999generalized}, \citet{deng1999minimum}), which are commonly used to evaluate competing screening designs. Finally, we identified the designs that perform best in terms of these two criteria.}

In so doing, we fill some of the remaining gaps in the literature on D- and A-optimal designs for main-effects models. Our first contribution is that we are the first authors to generate complete sets of non-isomorphic DA designs for numbers of runs that are not multiples of 4. Our work therefore builds a bridge between research on combinatorially constructed screening designs (focusing on numbers of runs that are powers of 2 or multiples of 4) and research on optimal screening designs (intended to deal also with numbers of runs that are not powers of 2 or multiples of 4). The latter research generally involves the use of heuristic algorithms, such as point-exchange and coordinate-exchange algorithms. These algorithms are known to fail to find optimal designs in many cases \citep[see, for example][]{Daniel}. Our second contribution is that our complete enumerations of non-isomorphic DA designs allow us to identify the DA designs that perform best in terms of the G- and G$_2$-aberration criteria for run sizes that are one and two more than a multiple of 4. These DA designs outperform those available in design of experiments software packages in terms of aliasing between main effects and interaction effects and among interaction effects.

% In Section~\ref{poc}, we provide two examples of DA designs that outperform the best known designs in the literature. 
In Section~\ref{sec:literature}, we review existing methods to construct DA designs for run sizes that are one and two more than a multiple of 4. These methods form the basis for our algorithms to construct our complete sets of non-isomorphic DA designs for these run sizes. In Section \ref{sec:alg_2mod4}, we present a design construction algorithm for run sizes that are two more than a multiple of four, and in Section \ref{sec:alg_1mod4}, we explain how this construction algorithm can be simplified for run sizes that are one more than a multiple of four. In Section \ref{sec:results}, we present the numbers of non-isomorphic DA designs for run sizes up to 18 enumerated with these algorithms, and report the numbers of DA designs that cannot be obtained from an orthogonal array. In Section \ref{computing_times}, we discuss the computational times required to generate our sets of DA designs. In Section \ref{sec:characterize}, we define the design selection criteria we used to evaluate our DA designs. To illustrate the benefit of complete design collections, we present in Section \ref{sec:best_designs} the best 17-run and 18-run DA designs we obtained and compare them to those obtained by ZM and HM. Finally, we summarize our findings in Section~\ref{sec:conclusion}.

\spacingset{1}
\section{Literature review} \label{sec:literature}
\spacingset{1.9}

% The two proof-of-concept examples demonstrate that DA designs that do not result from appending rows to an existing orthogonal array might be superior to those that can be derived from orthogonal arrays in terms of aliasing between main effects and interaction effects and among interaction effects. The challenge is to generate complete sets of non-isomorphic DA designs, so that we can identify those that minimize the aliasing between main effects and interactions and among interaction effects. 
Our algorithms to generate complete sets of non-isomorphic DA designs for numbers of runs that are one or two more than a multiple of 4 are inspired by theoretical results concerning the optimal forms of the information matrices for these numbers of runs, which we review in this section.

\spacingset{1}
\subsection{Run sizes that are one more than a multiple of four}
\spacingset{1.9}

Let us denote any run size that is one more than a multiple of 4 by $N_1$. \citet{cheng1980optimality} showed that any design involving $N_1$ runs and $k$ factors is a DA design if its $(k+1)$-dimensional information matrix is of the form
\begin{align} \label{form_n1}
\mathbf{X'X} = (N_1-1)\mathbf{I}_{k+1} + \mathbf{J}_{k+1},
\end{align}

\noindent where $\mathbf{I}_{k+1}$ is the $(k+1)$-dimensional identity matrix and $\mathbf{J}_{k+1}$ is a $(k+1) \times (k+1)$ {matrix with all entries equal to 1}. Any design that is isomorphic to a design with an information matrix of this form is {also a DA design.} A DA design with the information matrix in Equation~(\ref{form_n1}) has correlations of $1/(N_1+1)$ between any pair of factor columns. 
% This is visible in the top left parts of the color maps in Figure~\ref{fig:corr_17_11} of the DA designs in Tables~\ref{des:ZM_17_11} and~\ref{des:us_17_11}, where the light gray shade of the off-diagonal elements corresponds to a correlation of $1/(N_1+1)=1/18$.

An $N_1$-run DA design can easily be constructed in all unsaturated cases, for which $N_1 > k+1$. As a matter of fact, \citet{cheng1980optimality} pointed out that adding a row with any combination of $+1$s and $-$1s to an orthogonal array results in a DA design with one run more than the orthogonal array. This inspired ZM to study all $N_1$-run DA designs that can be obtained by adding a row of $+1$s and $-$1s to all orthogonal arrays in complete sets of non-isomorphic orthogonal arrays of run sizes 12 and 16. However, hitherto, no research has been performed to study whether DA designs constructed by adding one row to an orthogonal array are the only existing ones and whether these designs can be improved in terms of aliasing between main effects and interactions and among interactions. This is one of the gaps in the literature that we fill with the present paper.

For saturated cases, where $N_1 = k+1$, the construction based on orthogonal arrays does not work, because such arrays do not exist when the run size equals the number of factors. \citet{raghavarao1959some} showed that, when $N_1 = k + 1$, a DA design with an information matrix of the form given in Equation~(\ref{form_n1}) only exists when $\sqrt{2N_{1}-1}$ is an odd integer. Among all run sizes that are one more than a multiple of 4 and smaller than 100, those equaling 5, 13, 25, 41, 61 and 85 are the only ones for which this technical requirement is met. \citet{raghavarao1959some} also described one specific construction method for a DA design for each of these run sizes, based on the incidence matrix of a balanced incomplete block design. At present, it is unknown whether the DA designs produced by the method of \citet{raghavarao1959some} are the only existing ones and whether better designs exist in terms of aliasing between main effects and interactions and among interactions.

{For saturated cases where $\sqrt{2N_{1}-1}$ is not an odd integer, there are no optimal designs with information matrix~(\ref{form_n1}), and D-optimal designs can be different from A-optimal designs. We are not aware of any construction of A-optimal designs for these cases. However, truly D-optimal designs have been identified for run sizes 9, 17, 21, and 37 \citep{king2020direct}. }

\spacingset{1}
\subsection{Run sizes that are two more than a multiple of four}
\spacingset{1.9}

Let us denote any run size that is two more than a multiple of 4 by $N_2$. \citet{jacroux1983optimality} showed that any design involving $N_2$ runs and $k$ factors is a DA design if its $(k+1)$-dimensional information matrix is of the form
    \begin{align} \label{form_n2}
    \Gamma_{(i,j)} =
    \begin{bmatrix}
        \mathbf{A}_i & \mathbf{0}_{i \times j} \\
        \ \mathbf{0}_{j \times i} & \mathbf{A}_j
    \end{bmatrix},
    \end{align}
where $i=j=(k+1)/2$ in the event $k$ is odd, $i=k/2$ and $j=k/2+1$ or vice versa in the event $k$ is even, $
\mathbf{A}_i = (N_{2}-2)\mathbf{I}_{i} + 2\mathbf{J}_{i}
$ and $\mathbf{A}_j = (N_{2}-2)\mathbf{I}_{j} + 2\mathbf{J}_{j}.$ The difference between $\mathbf{A}_i$ and $\mathbf{A}_j$ is that $\mathbf{A}_i$ involves $i$ rows and columns and $\mathbf{A}_j$ involves $j$ rows and columns. Obviously, regardless of whether $k$ is even or odd, $i+j=k+1$. Any design with $N_2$ runs that is isomorphic to a design with an information matrix equal to that in Equation~(\ref{form_n2}) is also a DA design.

%When such a design exists, \citet{jacroux1983optimality} also confirm that the associated information matrix given in Equation (\ref{form_n2}) is uniquely optimal with respect to the D- and A-optimality criteria.

When $k$ is odd, $i=j=(k+1)/2$ and the optimal form of the information matrix is $\Gamma_{((k+1)/2,(k+1)/2)}$. DA designs with an information matrix of this form have $(k-1)/2$ factor columns that add up to 2, $(k+1)/2$ factor columns that add up to 0, and $(k^2-1)/4$ pairs of factor columns that are orthogonal to each other.

When $k$ is even, we assume without loss of generality that the first column of the main-effects model matrix corresponds with the intercept. There are two optimal forms of the information matrix:
\begin{itemize}[noitemsep,topsep=0pt]
\item The first optimal form has $i=k/2$ and $j=k/2+1$, and can therefore be denoted by $\Gamma_{(k/2,k/2+1)}$. DA designs with this optimal form involve $k/2-1$ factor columns that add up to 2 and $k/2+1$ factor columns that add up to 0. There are a total of $k^{2}/4 - 1$ pairs of factor columns that are orthogonal to each other.
\item The second optimal form has $i=k/2+1$ and $j=k/2$, and can therefore be denoted by $\Gamma_{(k/2+1,k/2)}$. DA designs with this optimal form possess $k/2$ factor columns that add up to 2 and $k/2$ factor columns that add up to 0. In these DA designs, $k^2/4$ pairs of factor columns are orthogonal to each other.
\end{itemize}
For all $N_2$-run DA designs with an information matrix of the form given in Equation(\ref{form_n2}), the pairs of columns that are orthogonal and thus have zero correlation involve one column that sums to 0 and one that sums to 2. Pairs of factor columns that add up to 2 have correlations of $2/(N_2+2)$ and pairs of factor columns that add up to 0 have correlations of $2/N_2$.

\citet{payne1974maximizing} explained that DA designs with $N_2$ runs and $k$ factors can be constructed by taking a normalized Hadamard matrix of order $N_{2}-2$, retaining any $k+1$ columns (including the first column that consists of only plus ones for the intercept) and adding two new rows. One of these two rows has all entries equal to $+1$ and the other row has a $+1$ as its first entry and is as balanced as possible in the numbers of $+1$ and $-1$ entries. \citet{galil1980d} pointed out that the original $(N_{2}-2)$-run design with $k$ factors need not be obtained by dropping columns from a Hadamard matrix. It suffices that it is an $(N_{2}-2)$-run orthogonal array. In addition, \citet{king2020direct} pointed out that the two extra rows can be any two rows with $-1$s and $+1$s, as long as they are as mutually orthogonal as possible. The construction methods of \citet{payne1974maximizing} and \citet{galil1980d} only work when $N_{2} > k+2$, since it requires an orthogonal array with $N_{2}-2$ rows and such orthogonal arrays only exist when $N_{2}-2 > k$.

\citet{Ehlich1964} showed that, in the saturated case, where {$k=N_{2}-1$}, a DA design with the information matrix given in Equation (\ref{form_n2}) only exists when $2N_{2}-2$ is a sum of two squared integers. For cases where this condition is satisfied, \citet{chadjipantelis1985supplementary} provided a construction based on circulant matrices. When {$k=N_{2}-2$, dropping any column from a DA design with $k=N_{2}-1$} (if known) produces a DA design \citep{galil1980d}.

For all saturated cases where $2N_{2}-2$ is not a sum of two squared integers, \citet{lin2016constructing} provided a construction for designs with information matrices which have a determinant very close to that of the determinant of the information matrix given in Equation~(\ref{form_n2}). \citet{chasiotis2018d} proved that the 22-run design obtained by \citet{lin2016constructing} is D-optimal and compared it with another design which is also D-optimal but has an information matrix different from the one of the design of \citet{lin2016constructing}. This implies that more than one form of the information matrix can be associated with a saturated D-optimal design when $2N_{2}-2$ is not a sum of two squared integers. However, it is unknown whether these designs are also A-optimal. For situations where $N_2 = k+2$ and $2N_{2}-2$ is not a sum of two squares, we are not aware of any work on the construction of DA designs with an information matrix of the form given in Equation (\ref{form_n2}).

For run sizes $N_2 \leq 18$, HM performed a systematic study of all non-isomorphic DA designs that can be obtained using the constructions of \citet{payne1974maximizing} and \citet{galil1980d}. However, there is no guarantee that these constructions yield a complete set of non-isomorphic $N_2$-run DA designs. Similarly, there is no guarantee that the construction of \citet{chadjipantelis1985supplementary} yields a complete set of non-isomorphic saturated DA designs when $2N_{2}-2$ is a sum of two squared integers. These are two other gaps in the literature that we address with this paper.

\spacingset{1}
\subsection{Conclusion}
\spacingset{1.9}

Except for some saturated cases, the optimal form of the information matrix for a DA design is known for run sizes $N_1$ and $N_2$. As a result, the information to generate all $N_1$- and $N_2$-run non-isomorphic DA designs is available. For DA designs with information matrices as given in Equations (\ref{form_n1}) or (\ref{form_n2}), assuming that the first column of the designs' model matrix is the intercept column, the first row and column in the information matrix informs us what the sum of each factor column in the DA design must be, and all off-diagonal values other than those in the first row and column give us the possible values for the inner products between any two factor columns. Using this information, we propose two new design construction methods, one for run sizes $N_1$ that are one more than a multiple of four and one for run sizes $N_2$ that are two more than a multiple of four.

\spacingset{1}
\section{Algorithm for enumerating DA designs with run sizes two more than a multiple of four} \label{sec:alg_2mod4}
\spacingset{1.9}

% Our enumeration procedure for DA designs with run sizes $N_2$ follows a multi-stage column extension procedure which is more complex than the one for run size $N_1$, due to the existence of two different optimal forms of the information matrix in the event the number of factors $k$ is even. 

Our enumeration procedure for DA designs with run sizes $N_2$ follows a multi-stage column extension procedure similar to the approach of \citet{schoen2010complete}. In Section~\ref{ssec:2mod4_cc}, we detail the column extension procedure. In Section \ref{ssec:2mod4_start}, we discuss the starting design of the enumeration. Section~\ref{ssec:2mod4_setcolumns} details the set of columns we consider for extension at each column extension stage, and finally, in Section \ref{ssec:2mod4_esteps}, we summarize all steps of our enumeration procedure.

\spacingset{1}
\subsection{Column-by-column approach} \label{ssec:2mod4_cc}
\spacingset{1.9}

From Equation (\ref{form_n2}), it is clear that for run size $N_2$, the optimal form for the information matrix for $k-1$ factors is a submatrix of the optimal form for the information matrix for $k$ factors, for any given value of $k$. Therefore, for run size $N_2$, a complete set of non-isomorphic DA designs with $k$ factors can be enumerated by extending the single non-isomorphic one-factor DA design, column by column, where the procedure of adding one column at a time is repeated until we obtain all non-isomorphic DA designs for the desired number of factors $k$. Due to the fact that there are two optimal forms of the information matrix for even values of $k$, the extension of a design with an odd number of factors differs from the extension of a design with an even number of factors. Another complication is that $N_2$-run DA designs with more than one factor possess factor columns that sum to 0 and factor columns that sum to 2. For this reason, we need to utilize two sets of candidate columns in the extension step of the enumeration algorithm, one with columns that sum to 0 and the other with columns that sum to 2.

\spacingset{1}
\subsubsection{Extending a design with an odd number of factors}
\spacingset{1.9}

There is only one optimal form of the information matrix for an $h$-factor DA design when $h$ is odd. Since $h+1$ is even in the event $h$ is odd, an $(h+1)$-factor DA design can have one of two optimal forms. Our extension step therefore has to generate designs with information matrices of both optimal forms $\Gamma_{((h+1)/2,(h+1)/2+1)}$ and $\Gamma_{((h+1)/2+1,(h+1)/2)}$ by extending DA designs of the form $\Gamma_{((h+1)/2,(h+1)/2)}$:
\begin{itemize}[noitemsep,topsep=0pt]
 \item Dropping any one of the factor columns that sum to 0 from any given $(h+1)$-factor DA design with an information matrix of the form $\Gamma_{((h+1)/2,(h+1)/2+1)}$ must yield a DA design with $h$ factors, which has an information matrix of the form $\Gamma_{((h+1)/2,(h+1)/2)}$. Therefore, any DA design with $h+1$ factors of the form $\Gamma_{((h+1)/2,(h+1)/2+1)}$ can only be created by adding a factor column that sums to 0 to a DA design of the form $\Gamma_{((h+1)/2,(h+1)/2)}$, where the added column has (i) inner products of 2 with all factor columns of the $h$-factor DA design that sum to 0 and (ii) inner products of 0 with all factor columns of the $h$-factor DA design that sum to 2.
\item For enumerating DA designs with information matrices of the form $\Gamma_{((h+1)/2+1,(h+1)/2)}$, it is useful to observe that dropping any one of the factor columns that sum to 2 from any given $(h+1)$-factor DA design with an information matrix of the form $\Gamma_{((h+1)/2+1,(h+1)/2)}$  yields an $h$-factor DA design with an information matrix of the form $\Gamma_{((h+1)/2,(h+1)/2)}$. Therefore, any DA design with $h+1$ factors with an information matrix of the form $\Gamma_{((h+1)/2+1,(h+1)/2)}$ can be constructed by adding a factor column that sums to 2 to an $h$-factor DA design with an information matrix of the form $\Gamma_{((h+1)/2,(h+1)/2)}$, where the added column has (i) inner products of 2 with all factor columns of the $h$-factor DA design that sum to 2 and (ii) inner products of 0 with all factor columns of the $h$-factor DA design that sum to 0.
\end{itemize}

\spacingset{1}
\subsubsection{Extending a design with an even number of factors}\label{kiekeboe}
\spacingset{1.9}

As $h+1$ is odd in the event $h$ is even and as there is only one optimal form for the information matrix of DA designs with an odd number of factors, there are two ways in which we can construct a complete set of all non-isomorphic $(h+1)$-factor DA designs when $h$ is even:
\begin{itemize}[noitemsep,topsep=0pt]
\item Any $(h+1)$-factor DA design can be constructed by adding a factor column that sums to 2 to an $h$-factor design with an information matrix of the optimal form $\Gamma_{(h/2,h/2+1)}$, where the added column has (i) inner products of 2 with all factor columns of the $h$-factor DA design that sum to 2 and (ii) inner products of 0 with all factor columns of the $h$-factor DA design that sum to 0. This is because dropping a column that sums to 2 from a $(h+1)$-factor DA design yields a DA design with an information matrix of the optimal form $\Gamma_{(h/2,h/2+1)}$.
\item Any $(h+1)$-factor DA design can also be constructed by adding a factor column that sums to 0 to an $h$-factor design with an information matrix of the optimal form $\Gamma_{(h/2+1,h/2)}$, where the added column has (i) inner products of 2 with all factor columns of the $h$-factor DA design that sum to 0 and (ii) inner products of 0 with all factor columns of the $h$-factor DA design that sum to 2. This is because dropping a column that sums to 0 from a $(h+1)$-factor DA design yields a DA design with an information matrix of the optimal form $\Gamma_{(h/2+1,h/2)}$.
\end{itemize}
As an $(h+1)$-factor DA design can be obtained from an $h$-factor design of either of the optimal forms, it suffices to use only one such form to construct a compete set of all non-isomorphic $(h+1)$-factor designs.

\spacingset{1}
\subsection{Starting design} \label{ssec:2mod4_start}
\spacingset{1.9}

It is convenient to start the enumeration with a two-factor design instead of a design with a single factor. We have two options for the two-factor starting design in the enumeration procedure for DA designs with $N_2$ runs: $\Gamma_{(1,2)}$ and $\Gamma_{(2,1)}$. It can be verified that the only $N_2$-run two-factor design with the former information matrix involves $(N_2+2)/4$ replicates of the design points $(-1,-1)$ and $(1,1)$ and $(N_2-2)/4$ replicates of the design points $(1,-1)$ and $(-1,1)$, and that the only $N_2$-run two-factor design with the latter information matrix involves $(N_2+2)/4$ replicates of the design points $(1,-1)$ and $(1,1)$ and $(N_2-2)/4$ replicates of the design points $(-1,-1)$ and $(-1,1)$. 
% This can be seen by setting up a system of equations whose unknowns are the numbers of replicates of the four design points, stating that (i) for $\Gamma_{(1,2)}$, both columns of the design should sum to 0 and that their inner product should be equal to 2, and (ii) for $\Gamma_{(2,1)}$, one column should sum to 2 and the other should sum to 0 and that their inner product should be equal to 0. Since the system of equations only has one solution for either of the forms, there is only one non-isomorphic two-factor DA design with $N_2$ runs for each optimal form: $\Gamma_{(1,2)}$ and $\Gamma_{(2,1)}$.

As explained in Section~\ref{kiekeboe}, it suffices to use only one of the two non-isomorphic two-factor designs to construct a complete set of all non-isomorphic three-factor designs. We arbitrarily chose the two-factor design with an information matrix of the form $\Gamma_{(2,1)}$ as the starting design for our enumeration for each given run size $N_2$.

\spacingset{1}
\subsection{Possible sets of columns for extension} \label{ssec:2mod4_setcolumns}
\spacingset{1.9}

In $N_2$-run DA designs with at least three factors, some factor columns sum to 2 and some factor columns sum to 0. For a given run size $N_2$, there are $N_{2} \choose (N_2+2)/2$ columns that sum to 2 and $N_{2} \choose N_{2}/2$ columns that sum to 0. We refer to the set of columns that sum to 2 by $\zeta_{N_2}^2$ and to the set of columns that sum to 0 by $\zeta_{N_2}^0$. So, any $N_2$-run design with three or more factors includes at least one column from the set $\zeta_{N_2}^2$ and at least one column from the set $\zeta_{N_2}^0$.

Since we start our enumeration from the two-factor DA design with the information matrix $\Gamma_{(2,1)}$ and given the optimal form of the information matrix of any DA design with three or more factors, not every column from the sets $\zeta_{N_2}^2$ and $\zeta_{N_2}^0$ is eligible for addition to the starting design. More specifically, due to the structure of the optimal form and our choice of the starting design, the only eligible columns from $\zeta_{N_2}^2$ are those that have an inner product of 2 with the first factor of the two-factor starting design and an inner product of 0 with the second factor of that design. Similarly, the only eligible columns from $\zeta_{N_2}^0$ are those that have an inner product of 2 with the second factor of the two-factor starting design and an inner product of 0 with the first factor of that design. This allows us to reduce the set of candidate columns substantially. We refer to the subsets of $\zeta_{N_2}^2$ and $\zeta_{N_2}^0$ containing only the eligible columns as $\zeta_{N_2}^{2*}$ and $\zeta_{N_2}^{0*}$, respectively. Only the columns in the sets $\zeta_{N_2}^{2*}$ and $\zeta_{N_2}^{0*}$ need to be considered for addition in our enumeration algorithm.

\spacingset{1}
\subsection{Enumeration steps} \label{ssec:2mod4_esteps}
\spacingset{1.9}

For a given run size $N_2$, our enumeration proceeds as follows:

\begin{itemize}[noitemsep,topsep=0pt]
    \item \textbf{Step 1: Generate the two-factor starting design}
    \item \textbf{Step 2: Generate the two sets of candidate columns for extension}
    \begin{itemize}[noitemsep,topsep=0pt]
        \item \textbf{Step 2a:} Generate the set $\zeta_{N_2}^2$ containing all possible columns that sum to 2. Then, reduce this set by only keeping columns that have an inner product of 2 with the first factor column of the starting design from Step 1 and an inner product of 0 with the second factor column of that design. This produces the first set of candidate columns, $\zeta_{N_2}^{2*}$, to be considered in Step 3.
        \item \textbf{Step 2b:} Generate the set $\zeta_{N_2}^0$ containing all possible columns that sum to 0. Then, reduce this set by only keeping columns that have an inner product of 0 with the first factor column of the starting design from Step 1 and an inner product of 2 with the second factor column of that design. This produces the second set of candidate columns, $\zeta_{N_2}^{0*}$, to be considered in Step 3.
    \end{itemize}
    \item \textbf{Step 3: Design extension}\\
    The enumeration for run sizes $N_2$ is done by extending the two-factor DA design constructed in Step 1 column by column until the desired number of factors $k$ has been reached. The design extension step for $N_2$ thus consists of $k-2$ stages, which we name $3, 4, \dots, k$. In a given stage $h$ of the extension step, all non-isomorphic $(h-1)$-factor DA designs are extended using one or both sets of candidate columns $\zeta_{N_2}^{2*}$ and $\zeta_{N_2}^{0*}$, depending on whether $h$ is odd or even. After each successful extension at stage $h$, the $h$-factor design obtained is converted into the Nauty minimal form \citep{mckay2014practical} using the OA package \citep{eendebak2019oapackage}. This conversion produces a unique design form that represents its isomorphism class. This unique representation, which we refer to as the design's canonical form, allows us to keep one design from each isomorphism class. These canonical forms may have an initial pair of columns that are different from the columns of the two-factor starting design. This is due to the conversion performed by Nauty to check the possible isomorphism of all designs generated.

Two databases are therefore utilized at each given stage $h$. The first database, $D_1$, stores all unique canonical forms of the non-isomorphic DA designs, while the second database, $D_2$, stores the corresponding designs that do have information matrices of the optimal form in Equation~(\ref{form_n2}). To extend a given $N_2$-run $(h-1)$-factor DA design with one additional factor, the enumeration will follow one or both of the following steps:
		\begin{itemize}[noitemsep,topsep=0pt]
        \item \textbf{Step 3.1:} Extension with a factor column from $\zeta_{N_2}^{0*}$
        \begin{itemize}[noitemsep,topsep=0pt]
            \item \textbf{Step 3.1.1:} Check whether the next candidate column from $\zeta_{N_2}^{0*}$ has an inner product equal to 0 with all factor columns that sum to 2 and an inner product equal to 2 with all factor columns that sum to 0 in the $(h-1)$-factor design being extended. If yes, continue to Step 3.1.2. Otherwise, repeat Step 3.1.1 considering the next candidate column from $\zeta_{N_2}^{0*}$.
            \item \textbf{Step 3.1.2:} Convert the resulting $h$-factor design into its canonical form. If the canonical form already exists in the first database $D_1$, then move to the next candidate column from $\zeta_{N_2}^{0*}$ and return to Step 3.1.1. Otherwise, add the new canonical form to the first database $D_1$, and the original extended design to the second database $D_2$.
       \end{itemize}
        \item \textbf{Step 3.2:} Extension with a factor column from $\zeta_{N_2}^{2*}$
        \begin{itemize}[noitemsep,topsep=0pt]
            \item \textbf{Step 3.2.1:} Check whether the next candidate column from $\zeta_{N_2}^{2*}$ has an inner product equal to 2 with all factor columns that sum to 2 and an inner product equal to 0 with all factor columns that sum to 0 in the $(h-1)$-factor design being extended. If yes, continue to Step 3.2.2. Otherwise, repeat Step 3.2.1 considering the next candidate column from $\zeta_{N_2}^{2*}$.
            \item \textbf{Step 3.2.2:} Convert the resulting $h$-factor design into its canonical form. If the canonical form already exists in the first database $D_1$, then move to the next candidate column from $\zeta_{N_2}^{2*}$ and return to Step 3.2.1. Otherwise, add the new canonical form to the first database $D_1$, and the original extended design to the second database $D_2$.
        \end{itemize}
    \end{itemize}
\end{itemize}
		
The procedure begins by extending the two-factor DA design with the information matrix $\Gamma_{(2,1)}$ for a given run size $N_2$ to generate all non-isomorphic three-factor DA designs by adding factor columns from the set $\zeta_{N_2}^{0*}$ (Step 3.1). This produces all three-factor designs with information matrix $\Gamma_{(2,2)}$. The next step is to create all four-factor DA designs. The information matrices of these designs can be of two forms, namely $\Gamma_{(3,2)}$ and $\Gamma_{(2,3)}$. To obtain all four-factor designs with an information matrix of the form $\Gamma_{(3,2)}$, all three-factor designs are extended by adding new factor columns from the set $\zeta_{N_2}^{2*}$ (Step 3.2). To obtain all four-factor designs with an information matrix of the form $\Gamma_{(2,3)}$, all three-factor designs are extended by adding new factor columns from the set $\zeta_{N_2}^{0*}$ (Step 3.1). To obtain all non-isomorphic DA designs with five factors, all of which have an information matrix of the form $\Gamma_{(3,3)}$, we can either extend four-factor designs with an information matrix of the form $\Gamma_{(3,2)}$ by adding a factor from the set $\zeta_{N_2}^{0*}$ (Step 3.1) or extend four-factor designs with an information matrix of the form $\Gamma_{(2,3)}$ by adding a factor from the set $\zeta_{N_2}^{2*}$ (Step 3.2). Both of these approaches lead to the same set of designs. We chose to extend the designs of the form $\Gamma_{(2,3)}$, by adding new factor columns from the set $\zeta_{N_2}^{2*}$ (Step 3.2). In general, every time we are presented with this choice, we extend the designs with an information matrix of the form $\Gamma_{(\lambda,\lambda+1)}$ by adding factor columns from the set $\zeta_{N_2}^{2*}$ (Step 3.2). This is because we found that for all run sizes $N_2$ that we study in this paper, the set $\zeta_{N_2}^{2*}$ has fewer columns than set $\zeta_{N_2}^{0*}$. After obtaining all DA designs with an information matrix of the form $\Gamma_{(3,3)}$, we need to produce designs of with information matrices of the forms $\Gamma_{(4,3)}$ and $\Gamma_{(3,4)}$. This cycle continues until all DA designs for the desired number of factors $k$ are enumerated.

\spacingset{1}
\section{Algorithm for enumerating DA designs with run sizes one more than a multiple of four} \label{sec:alg_1mod4}
\spacingset{1.9}

The algorithm to enumerate DA designs for run size $N_1$ follows a column-by-column extension procedure similar to the algorithm for run size $N_2$. However, it is simpler because we have only one optimal form of the information matrix for both odd and even $k$, one starting design and one set of candidate columns $\zeta_{N_1}^*$. The same enumerations steps as those given in Section \ref{ssec:2mod4_esteps} apply with minor changes. The two-factor starting design in Step 1 has the information matrix $(N_1-1)\mathbf{I}_{3} + \mathbf{J}_{3}$, $(N_1-1)/4$ replicates of the design points $(-1,-1)$, $(-1,1)$ and $(1,-1)$, and $(N_1+3)/4$ replicates of the design point $(1,1)$. The appropriate candidate set $\zeta_{N_1}^*$ created in Step 2 consists of all columns that sum to $1$ and that have an inner product equal to $1$ with both columns of the starting design. Finally, for Step 3, we have two substeps. In Step 3.1, we check whether the candidate column has inner products equal to 1 with all factor columns of the design that is extended. If yes, we move to Step 3.2, where we convert the newly created design into its canonical form and store the original extended design in database $D_2$ if its canonical form does not exist yet in database $D_1$.
 
\spacingset{1}
\section{Numbers of non-isomorphic DA designs}  \label{sec:results}
\spacingset{1.9}

In this section, we present the numbers of non-isomorphic DA designs for all run sizes $N_1$ and $N_2$ up to 18. For each of the run sizes, we also list the numbers of non-isomorphic DA designs that cannot be obtained by extending an orthogonal array and which were therefore not considered by ZM and HM.
%We do this by checking whether dropping any row or any set of two rows from the designs we obtained produces an orthogonal array. The results for run sizes $N_1$ are presented in Section \ref{sec:results_1}, while those for run sizes $N_2$ are presented in Section \ref{sec:results_2}.

\spacingset{1}
\subsection{Run size is one more than a multiple of four} \label{sec:results_1}
\spacingset{1.9}

Table~\ref{tab:no_designs_1mod4} shows the numbers of non-isomorphic DA designs for run sizes $N_1$ equal to 5, 9, 13 and 17, and up to 15 factors. We do not report results concerning 9-run 8-factor designs and 17-run 16-factor designs, since these are saturated cases where $\sqrt{2N_{1}-1}$ is not an odd integer and for which there is no design with an information matrix of the form in Equation~(\ref{form_n1}). Therefore, our enumeration procedure cannot produce DA designs for these two cases.

For each run size $N_1$, there are two columns in the table. The first column, labeled \emph{T}, gives the total number of non-isomorphic DA designs. The second column, labeled \emph{T\textsubscript{no}}, gives the number of designs that cannot be obtained from orthogonal arrays and which were therefore not considered by ZM. For certain combinations of the run size $N_1$ and the number of factors $k$, a significant number of DA designs cannot be obtained by appending a row to an orthogonal array. For instance, more than one third of the 13-run 6-factor designs and of the 17-run 8-factor designs cannot be obtained this way. The enumeration of these designs is one major contribution of this paper.

\spacingset{1}
\begin{table}[!t]
  \centering
  \caption{Total numbers of $k$-factor DA designs (\emph{T}) and numbers of DA designs that cannot be obtained from orthogonal arrays (\emph{T\textsubscript{no}}) for run sizes $N_1$ equaling 5, 9, 13 and 17.}
    \begin{tabular}{cccrccrrcrrr}
    \toprule
$k$          & \multicolumn{2}{c}{$N_1=5$} &       & \multicolumn{2}{c}{$N_1=9$} &       & \multicolumn{2}{c}{$N_1=13$} &       & \multicolumn{2}{c}{$N_1=17$} \\
\cmidrule{2-12}    & \emph{T} & \emph{T\textsubscript{no}} &  & \emph{T} & \emph{T\textsubscript{no}} & & \emph{T} & \emph{T\textsubscript{no}} & & \emph{T} & \emph{T\textsubscript{no}} \\
    \midrule
    3     & 2     & 0     &       & 3     & 0     &       & 4     & 0     &       & 5     & 0 \\
    4     & 1     & 1     &       & 4     & 0     &       & 7     & 2     &       & 14    & 2 \\
    5     &       &       &       & 3     & 0     &       & 14    & 4     &       & 58    & 13 \\
    6     &       &       &       & 3     & 0     &       & 20    & 7     &       & 293   & 96 \\
    7     &       &       &       & 4     & 0     &       & 22    & 6     &       & 1224  & 465 \\
    8     &       &       &       & -     & -     &       & 23    & 5     &       & 3172  & 1082 \\
    9     &       &       &       &       &       &       & 17    & 4     &       & 5224  & 1261 \\
    10    &       &       &       &       &       &       & 10    & 2     &       & 6312  & 952 \\
    11    &       &       &       &       &       &       & 9     & 1     &       & 5844  & 488 \\
    12    &       &       &       &       &       &       & 1     & 1     &       & 4041  & 186 \\
    13    &       &       &       &       &       &       &       &       &       & 2017  & 52 \\
    14    &       &       &       &       &       &       &       &       &       & 752   & 12 \\
    15    &       &       &       &       &       &       &       &       &       & 227   & 3 \\
    16    &       &       &       &       &       &       &       &       &       & -     & - \\
    \bottomrule
    \end{tabular}%
  \label{tab:no_designs_1mod4}%
\end{table}%
\spacingset{1.9}

In the majority of cases, the number of non-isomorphic $N_1$-run DA designs we obtained exceeds the numbers of non-isomorphic orthogonal arrays with one run less and three runs more, as reported by \citet{schoen2010complete}. For instance, when $k=10$, there are 6312 non-isomorphic 17-run DA designs, while there are only 78 and 2389 non-isomorphic orthogonal arrays with 16 and 20 runs, respectively. This implies that obtaining complete sets of non-isomorphic DA designs for run sizes that are one more than a multiple of four is computationally more challenging than obtaining complete sets of non-isomorphic orthogonal arrays, which have a run size that is a multiple of four.

Confirming the findings of \citet{orrick2008enumeration}, we found only one non-isomorphic DA design for each of the saturated cases with 5 and 13 runs. So, there is only one 5-run 4-factor DA design and one 13-run 12-factor DA design and these cannot be obtained by appending a row to 4- and 12-run DA designs, because these designs have at most 3 and 11 factors, respectively.

\spacingset{1}
\subsection{Run size is two more than a multiple of four} \label{sec:results_2}
\spacingset{1.9}

Table~\ref{tab:no_designs_2mod4} shows the numbers of non-isomorphic DA designs for run sizes 6, 10, 14 and 18, and up to 17 factors. For even values of $k$, we provide the numbers of designs for each of the two optimal forms in separate rows. Otherwise, the layout of Table~\ref{tab:no_designs_2mod4} is the same as that of Table~\ref{tab:no_designs_1mod4}.

The numbers of DA designs for run sizes 10, 14 and 18 are substantially higher than those for run sizes 9, 13 and 17, respectively. Also, in many cases, the number of designs that cannot be obtained from orthogonal arrays exceeds the number that can be obtained from such designs. These cases are characterized by a \emph{T\textsubscript{no}} value that is larger than \emph{T}$/2$. This suggests that our enumeration has more added value for run sizes that are two more than a multiple of four than for run sizes that are one more. The enumeration of the $N_2$-run designs that cannot be obtained from orthogonal arrays is a second major contribution of this paper.

\spacingset{1}
\begin{table}[!t]
  \centering
  \caption{Total numbers of $k$-factor DA designs (\emph{T}) and numbers of DA designs that cannot be obtained from orthogonal arrays (\emph{T\textsubscript{no}}) for run sizes $N_2$ equaling 6, 10, 14 and 18.}
    \begin{tabular}{cccclcclrrlrr}
    \toprule
$k$     &Optimal form     & \multicolumn{2}{c}{$N_2=6$} &       & \multicolumn{2}{c}{$N_2=10$} &       & \multicolumn{2}{c}{$N_2=14$} &       & \multicolumn{2}{c}{$N_2=18$} \\
\cmidrule{3-13}    &  & \emph{T} & \emph{T\textsubscript{no}} &  & \emph{T} & \emph{T\textsubscript{no}} & & \emph{T} & \emph{T\textsubscript{no}} & & \emph{T} & \emph{T\textsubscript{no}} \\
    \midrule
    3   & $\Gamma_{2,2}$ & 2     & 0     &       & 3     & 0     &       & 4     & 0     &       & 5     & 0 \\
    4  & $\Gamma_{2,3}$ & 1     & 1     &       & 5     & 1     &       & 7     & 3     &       & 18    & 4 \\
    4  & $\Gamma_{3,2}$ & 1     & 1     &       & 6     & 1     &       & 9     & 3     &       & 24    & 5 \\
    5  & $\Gamma_{3,3}$  & 1     & 1     &       & 9     & 2     &       & 37    & 21    &       & 241   & 103 \\
    6  & $\Gamma_{3,4}$ &       &       &       & 11    & 4     &       & 108   & 66    &       & 2905  & 1792 \\
    6  & $\Gamma_{4,3}$ &       &       &       & 12    & 3     &       & 133   & 83    &       & 3730  & 2314 \\
    7  & $\Gamma_{4,4}$  &       &       &       & 16    & 6     &       & 295   & 171   &       & 40048 & 28951 \\
    8  & $\Gamma_{4,5}$ &       &       &       & 2     & 2     &       & 334   & 158   &       & 177887 & 127508 \\
    8  & $\Gamma_{5,4}$ &       &       &       & 4     & 4     &       & 436   & 205   &       & 222379 & 159459 \\
    9  & $\Gamma_{5,5}$  &       &       &       & 1     & 1     &       & 428   & 163   &       & 561182 & 346154 \\
    10 & $\Gamma_{5,6}$ &       &       &       &       &       &       & 273   & 91    &       & 856884 & 379866 \\
     10 & $\Gamma_{6,5}$ &       &       &       &       &       &       & 302   & 94    &       & 1022082 & 452174 \\
    11 & $\Gamma_{6,6}$  &       &       &       &       &       &       & 157   & 53    &       & 1383882 & 392406 \\
    12  & $\Gamma_{6,7}$ &       &       &       &       &       &       & 8     & 8     &       & 1362650 & 216854 \\
     12 & $\Gamma_{7,6}$ &       &       &       &       &       &       & 11    & 11    &       & 1589648 & 254386 \\
    13 & $\Gamma_{7,7}$  &       &       &       &       &       &       & 1     & 1     &       & 1262389 & 116622 \\
    14  & $\Gamma_{7,8}$ &       &       &       &       &       &       &       &       &       & 611099 & 32843 \\
     14 & $\Gamma_{8,7}$ &       &       &       &       &       &       &       &       &       & 696028 & 36280 \\
    15 & $\Gamma_{8,8}$  &       &       &       &       &       &       &       &       &       & 181844 & 8860 \\
    16  & $\Gamma_{8,9}$ &       &       &       &       &       &       &       &       &       & 77    & 77 \\
     16 & $\Gamma_{9,8}$ &       &       &       &       &       &       &       &       &       & 80    & 80 \\
    17 & $\Gamma_{9,9}$  &       &       &       &       &       &       &       &       &       & 4     & 4 \\
    \bottomrule
    \end{tabular}%
  \label{tab:no_designs_2mod4}%
\end{table}%
\spacingset{1.9}

Due to the existence of the two optimal forms for the information matrix for even values of $k$, the numbers of DA designs for even $k$ values are sometimes larger than those for the next higher values of $k$. This happens for 8-, 10- and 12-factor 14-run designs, and for 10-, 12-, 14- and 16-factor 18-run designs.

Except for the saturated cases, for any number of factors $k$, the numbers of 10-, 14- and 18-run DA designs are larger than the numbers of DA designs for the two nearest run sizes that are multiples of four, as reported by \citet{schoen2010complete}. For example, when $k=9$, there are 428 non-isomorphic 14-run DA designs, while there is only one non-isomorphic 12-run DA design and there are only 87 non-isomorphic 16-run DA designs. As a result, the enumeration of all non-isomorphic DA designs with run sizes that are two more than a multiple of four is computationally more challenging than for run sizes that are multiples of four.

For the four saturated cases in Table~\ref{tab:no_designs_2mod4}, \citet{COHN1994214} reported the numbers of non-isomorphic DA designs. For three of the four saturated cases, our results match his. However, he reported three non-isomorphic 18-run 17-factor DA designs, while we found four. This is yet another contribution of our work.

\spacingset{1}
\section{Computing times} \label{computing_times}
\spacingset{1.9}

All our $N_1$- and $N_2$-run designs were generated using an Intel\textsuperscript{®} Core\textsuperscript{TM} i9-10980XE CPU @ 3.00 GHz processor using a single core. The computing times required to generate the entire set of non-isomorphic DA designs for run sizes 5, 6, 9, 10 and 13 are all less than one second. The times required to generate all DA designs for run sizes 14, 17 and 18 and each value of the number of factors $k$ are given in Table~\ref{tab:compute_time}. The times reported in this table for given numbers of factors $k$ are the total times required to extend the two-factor starting designs until all $k$-factor designs were obtained. This implies that, for example, extending the set of all non-isomorphic 14-factor 17-run DA designs with an extra column requires $12.21 - 11.91=0.30$  seconds of computing time.

The computational time required to generate the 18-run designs thus was substantial, but not problematic. This is because any researcher only has to perform the enumeration once if the full catalog of non-isomorphic designs is stored properly. Moreover, the enumeration algorithm can be parallelized to reduce the computational time for larger numbers of runs.

\spacingset{1}
\begin{table}[!t]\caption{Computing times for designs with $k$ factors and run sizes $N$ equal to 14, 17 and 18.}\label{tab:compute_time}
\centering
\begin{tabular}{rrcrcrc}
\toprule
$k$	& 	\multicolumn{2}{c}{$N=14$}		&	\multicolumn{2}{c}{$N=17$}		&	\multicolumn{2}{c}{$N=18$}\\	
\midrule
3	& 	0.15 & sec & 0.55	&	sec		&	2.68	&	sec\\
4	& 	0.49 & sec & 1.96	&	sec		&	10.58	&	sec\\
5	& 	0.68 & sec & 3.49	&	sec		&	15.16	&	sec\\
6	& 	1.41 & sec & 6.73	&	sec		&	60.49	&	sec\\
7	& 	2.09 & sec & 17.36	&	sec		&	3.79	&	min\\
8	& 	5.12 & sec & 51.75	&	sec		&	65.13	&	min\\
9	& 	6.79 & sec & 2.18	&	min		&	3.14	&	hr\\
10	& 	10.60 & sec & 4.31	&	min		&	16.80	&	hr\\
11	& 	11.71 & sec & 6.92	&	min		&	1.15	&	day\\
12	& 	13.00 & sec & 9.38	&	min		&	2.63	&	day\\
13	& 	13.09 & sec & 11.08	&	min		&	3.51	&	day\\
14	& 	&&11.91	&	min	&	5.12	&	day\\
15	& 	&&12.21	&	min		&	5.80	&	day\\
16	& 		&	& 		&		&	6.19	&	day\\
17	& 		&		& 		&		&	6.19	&	day\\
\bottomrule
\end{tabular}
\end{table}
\spacingset{1.9}

\spacingset{1}
\section{Design characterization} \label{sec:characterize}
\spacingset{1.9}

Our complete enumeration of non-isomorphic DA designs with run sizes $N_1$ and $N_2$ allowed us to identify the designs that minimize the aliasing between main effects and two-factor interactions, and among the two-factor interactions. To this end, we utilized the G-aberration criterion of \citet{deng1999generalized} and the G$_2$-aberration criterion of \citet{deng1999minimum}. Section~\ref{ssec:g_ab_ch4} introduces the first of these criteria, while Section~\ref{ssec:g2_ab_ch4} discusses the second.

\spacingset{1}
\subsection{G-aberration} \label{ssec:g_ab_ch4}
\spacingset{1.9}

G-aberration is based on the so-called J\textsubscript{$s$} characteristics or J\textsubscript{$s$} values. The J\textsubscript{1} values of a design are the absolute sums of the elements in each of the main-effect contrast columns. All other J\textsubscript{$s$} values, for $s$ ranging from 2 to $k$, are absolute values of the sums of the elements in $s$-factor interaction contrast columns \citep{deng1999generalized, schoen2017two}. The most important among the J\textsubscript{$s$} values are the J\textsubscript{2}, J\textsubscript{3}  and J\textsubscript{4} values, which quantify the extent to which main effects are aliased with each other, main effects are aliased with two-factor interactions, and two-factor interactions are aliased with other two-factor interactions, respectively. For values of $s$ exceeding 4, the J\textsubscript{$s$} values are related to higher-order interactions. Since these are generally negligible, these J\textsubscript{$s$} values are rather unimportant for practical purposes.

For each value of $s$, there are $k \choose s$ different $s$-factor interaction contrast columns for which a J\textsubscript{s} value needs to be calculated. Consequently, there are $k \choose s$ different J\textsubscript{s} values for a given $s$ value. These can be summarized in a frequency vector F\textsubscript{$s$}, starting with the frequency of the largest J\textsubscript{s} value and ending with the frequency of the smallest J\textsubscript{$s$} value. The concatenation of the frequency vectors F\textsubscript{1}, F\textsubscript{2}, \dots, F$_k$ is called the confounding frequency vector. Designs that minimize the entries of the confounding frequency vector from left to right sequentially minimize the most severe aliasing among the factors' effects, namely the aliasing among the main effects, between the main effects and the two-factor interactions and among the two-factor interactions. Such designs are called minimum G-aberration designs. The design evaluation criterion based on the confounding frequency vector is called the G-aberration criterion. The criterion is in line with the hierarchy principle, which states that lower-order effects are more likely to be active than higher-order effects. Minimizing the aliasing among lower-order effects, such as main effects and two-factor interactions, therefore ought to be prioritized.

An added value of the G-aberration criterion is that it distinguishes between factor effects that are completely aliased, partially aliased and not aliased at all. In the event a J\textsubscript{$s$} value is equal to the run size, then the effects involved are completely aliased. In the event a J\textsubscript{$s$} value is zero, then the effects involved are not aliased at all. In intermediate cases, the effects involved are partially aliased, and larger J\textsubscript{$s$} values correspond to more severe aliasing. A minimum G-aberration design therefore sequentially minimizes the number of completely aliased effects, starting with the number of completely aliased main effects and the number of completely aliased two-factor interactions. 
% The performance of a design in terms of the G-aberration criterion can be visualized by a color map of absolute correlations between the contrast columns for the main effects and the interaction effects. This is because the J\textsubscript{$s$} values are proportional to the absolute correlations between the factor effects' contrast columns. Design that minimize the G-aberration criterion sequentially minimize the number of dark gray cells, the number of medium gray cells, etc. in the different parts of the color maps.

In the characterization of our complete sets of non-isomorphic DA designs, we ignored the frequency vectors F\textsubscript{5}, \dots, F\textsubscript{$k$}, because they are concerned with higher-order interaction effects. We also ignore the F\textsubscript{1} and F\textsubscript{2} vectors. For run sizes that are one more than a multiple of 4, this is because all DA designs have factor columns that sum to one and because the inner product of each pair of factor columns is one as well. Therefore, all $N_1$-run DA designs have the same F\textsubscript{1} and F\textsubscript{2} vectors. So, the smallest $s$ value for which the F\textsubscript{$s$} vectors of the $N_1$-run DA designs can differ is 3. For run sizes that are two more than a multiple of 4, we also ignore the F\textsubscript{1} and F\textsubscript{2} vectors. For $N_2$-run DA designs with odd numbers of factors $k$, this is justified by the fact that these designs do not differ with respect to these vectors. Likewise, none of the $N_2$-run DA designs with an even number of factors and a given optimal form of the information matrix differs with respect to the F\textsubscript{1} and F\textsubscript{2} vectors. It should be pointed out, however, that DA designs with an even number of factors and an information matrix of the form $\Gamma_{(k/2,k/2+1)}$ have $k/2-1$ factor columns that sum to two and $k/2+1$ factor columns that sum to zero, while DA designs with an even number of factors and an information matrix of the form $\Gamma_{(k/2+1,k/2)}$ have $k/2$ factor columns that sum to two and $k/2$ factor columns that sum to zero. Therefore, the former type of DA design with an even number of factors has a slightly better F\textsubscript{1} vector and is therefore, technically speaking, preferable in terms of the G-aberration criterion.

\spacingset{1}
\subsection{G$_2$-aberration} \label{ssec:g2_ab_ch4}
\spacingset{1.9}

The G$_2$-aberration criterion is based on overall measures of the extent to which effects of one order are aliased with effects of another order. These overall measures of aliasing are derived from alias matrices, quantifying the bias in main-effects estimates in the presence of active interaction effects.

Denote by $\mathbf{X}_m$ the model matrix for the main-effects model including the intercept. If there are no active interaction effects, then the expected value of the response $\mathbf{Y}$ under the main-effects model is $\mathbf{X}_m \bm{\beta}_m$, where the first entry in $\bm{\beta}_m$ is the intercept and the remaining $k$ entries correspond to the factors' main effects. When there are active interaction effects of order $i$, this expected value becomes $\mathbf{X}_m \bm{\beta}_m + \mathbf{X}_i\bm{\beta}_i,$ where $\mathbf{X}_i$ is the matrix with all $i$-th order interaction contrast columns and $\bm{\beta}_i$ is the vector with all $i$-th order interaction effects. The matrix $(\mathbf{X}_m'\mathbf{X}_m)^{-1}\mathbf{X}_m'\mathbf{X}_i$ is referred to as the alias matrix for the main-effects model associated with the $i$th-order interaction effects. We denote it by $\mathbf{A}_i$. The product $\mathbf{A}_i\bm{\beta}_i$ is the bias in the estimates of the intercept and the main effects, if we fit the main-effects model ignoring the $i$th-order interaction effects. Since we generally care less about the bias in the estimate of the intercept, we ignore the first row of $\mathbf{A}_i$, and denote the matrix obtained by dropping that row by $\mathbf{A}^{*}_{i}$.

To minimize the overall bias in the main effects due to the $i$th-order interaction effects, \citet{deng1999minimum} suggested to minimize $C_i = tr({\mathbf{A}^{*}_{i} \mathbf{A}^{*}_{i}}').$ The vector $(C_2, C_3, \dots, C_{k-1})$ quantifies the bias in the main-effect estimates in the presence of active interactions of orders 2, 3, \dots, $k-1$, respectively. Due to the hierarchy principle, lower-order interactions are more likely to be active than higher-order interactions. The minimization of $C_2$ should therefore be prioritized over the minimization of $C_3$, the minimization of $C_3$ should be prioritized over the minimization of $C_4$, etc. This reasoning inspired \citet{deng1999minimum} to propose the G$_2$-aberration criterion: a design that sequentially minimizes the vector $(C_2, C_3, \dots, C_{k-1})$ performs best in terms of the G$_2$-aberration criterion and is called the minimum G$_2$-aberration design. Such a design minimizes the overall bias in the estimates of the main effects in the presence of higher-order interaction effects.

The quantities $C_2$ and $C_3$ can also be interpreted as measures of the overall aliasing between main effects and two-factor interactions and the overall aliasing among the two-factor interactions, respectively. In a similar fashion, the remaining elements of the vector $(C_2, C_3, \dots, C_{k-1})$ can be interpreted as measures of aliasing of higher-order interaction effects. For this reason, we only consider the $C_2$ and $C_3$ values when characterizing the non-isomorphic DA designs we enumerated.

One drawback of the fact that the G$_2$-aberration criterion uses measures of overall aliasing rather than aliasing of individual effects is that minimum G$_2$-aberration designs generally involve more completely aliased effects than minimum G-aberration designs. However, both minimum G- and minimum G$_2$-aberration designs minimize the aliasing among the factors' effects in a meaningful way. We therefore refer to both kinds of designs as minimally aliased designs in this paper.

\spacingset{1}
\section{Minimally aliased DA designs} \label{sec:best_designs}
\spacingset{1.9}

To illustrate the benefit of complete catalogs of DA designs, we report the characteristics of the minimally aliased designs with 17 and 18 runs. We provide all the designs discussed in the supplementary material to this paper.

\spacingset{1}
\subsection{Minimally aliased DA designs with 17 runs} \label{ssec:bd_1mod4}
\spacingset{1.9}
% Our enumeration and characterization of DA designs for run sizes that are one more than a multiple of four allowed us to identify one 13-run design that outperforms those reported in the literature in terms of G-aberration and one 13-run design that outperforms those reported in the literature in terms of G\textsubscript{2}-aberration. 
Our enumeration for run sizes 17 runs yielded seven DA designs that outperform those recommended in the literature in terms of G-aberration. Among these seven designs, there are three minimum G\textsubscript{2}-aberration 17-run designs that perform better in terms of G-aberration than the minimum G\textsubscript{2}-aberration designs reported by ZM. 

Our work confirmed that all the 17-run designs they recommended minimize the G\textsubscript{2}-aberration. Generally, however, their 17-run designs can be improved upon in term of G-aberration. In four cases, the improved design we found cannot be constructed by adding a row to an orthogonal array.

In Table~\ref{tab:bd_1mod4}, we present the maximum J\textsubscript{3} and maximum J\textsubscript{4} values, their frequencies, and the $C_2$ and $C_3$ values for the 17-run minimum G- and minimum G\textsubscript{2}-aberration designs, and for the designs presented by ZM. We limit ourselves to cases with values of $k$ where $17 < 2^k$, because we believe that the designs for $17 > 2^k$ are less interesting from a practical point of view. In the event Table~\ref{tab:bd_1mod4} has only one row for a given value of $k$, the design of ZM is a minimum G- and a minimum G\textsubscript{2}-aberration design. When the minimum G-aberration design is different from the minimum G\textsubscript{2}-aberration design for a give value of $k$, the first row characterizes the minimum G-aberration design, while the second row characterizes the minimum G\textsubscript{2}-aberration design. In many cases, we found multiple non-isomorphic minimum G\textsubscript{2}-aberration designs. For these cases, Table~\ref{tab:bd_1mod4} shows the characteristics of the minimum G\textsubscript{2}-aberration design that performs best in terms of the G-aberration criterion. For cases where the design presented by ZM differs from the minimum G-aberration design and/or the minimum G\textsubscript{2}-aberration design, we use an additional row to show the characteristics of that design.

\spacingset{1}
\begin{table}[!t]
  \centering
  \caption{Properties the $k$-factor DA designs with minimum G-aberration (mG) and minimum G\textsubscript{2}-aberration (mG\textsubscript{2}) and of those presented by ZM for run size $N_1$ equal to 17. nOA: design cannot be derived from an OA; J$_3^m$: maximum J$_3$ value; J$_4^m$: maximum J$_4$ value; $f$: frequency.}
    \begin{tabular}{rccccccrr}
    \toprule
    $k$     & mG & mG\textsubscript{2} & ZM & nOA & (J$_3^m$, $f$) &  (J$_4^m$, $f$) & $C_2$    & $C_3$ \\
    \midrule
    5     & x     & x     & x     &       & (\phantom{0}1, 10)    & (\phantom{0}1, \phantom{00}5)     & 0.103 & 0.103 \\
     6     & x     & x     & x     &       & (\phantom{0}1, 20)    & (17, \phantom{00}1)     & 0.170 & 11.062 \\
    7     & x     & x     & x     &       & (\phantom{0}1, 35)    & (17, \phantom{00}3)     & 0.255 & 26.196 \\
      8     & x     & x     & x     &       & (\phantom{0}1, 56)    & (17, \phantom{00}6)     & 0.358 & 52.518 \\
       9     & x     &       &       & x     & (\phantom{0}7, 42)    & (17, \phantom{00}1)     & 21.655 & 37.338 \\
       9   &       & x     & x     &       & (15, \phantom{0}4)     & (17, \phantom{00}6)     & 11.112 & 53.314 \\
      10    & x     &       &       & x     & (\phantom{0}7, 60)    & (17, \phantom{00}3)     & 30.556 & 61.852 \\
    10   &       & x     &       &       & (\phantom{0}9, \phantom{0}8)     & (17, \phantom{0}10)    & 22.049 & 72.988 \\
     10    &       & x     & x     &       & (15, \phantom{0}8)     & (17, \phantom{0}18)    & 22.049 & 72.988 \\
     11    & x     &       &       & x     & (\phantom{0}9, \phantom{0}6)     & (17, \phantom{00}3)     & 36.568 & 98.882 \\
     11   &       & x     &       &       & (\phantom{0}9, 12)    & (17, \phantom{00}8)     & 33.144 & 105.361 \\
     11    &       & x     & x     &       & (15, 12)    & (17, \phantom{0}26)    & 33.144 & 105.361 \\
      12    & x     &       &       & x     & (\phantom{0}9, 10)    & (17, \phantom{00}6)     & 47.712 & 148.222 \\
      12    &       & x     &       &       & (\phantom{0}9, 16)    & (17, \phantom{0}15)    & 44.376 & 157.598 \\
      12   &       & x     & x     &       & (15, 16)    & (17, \phantom{0}39)    & 44.376 & 157.598 \\
      13    & x     & x     &       &       & (\phantom{0}9, 22)    & (17, \phantom{0}15)    & 61.187 & 221.907 \\
     13   &       & x     & x     &       & (15, 22)    & (17, \phantom{0}55)    & 61.187 & 221.907 \\
     14    & x     & x     &       &       & (\phantom{0}9, 28)    & (17, \phantom{0}21)    & 78.158 & 310.040 \\
     14    &       & x     & x     &       & (15, 28)    & (17, \phantom{0}77)    & 78.158 & 310.040 \\
     15    & x     & x     &       &       & (15, \phantom{0}7)     & (17, \phantom{0}21)    & 98.027 & 422.051 \\
     15   &       & x     & x     &       & (15, 35)    & (17, 105)   & 98.027 & 422.051 \\
    \bottomrule
    \end{tabular}%
  \label{tab:bd_1mod4}%
\end{table}%
\spacingset{1.9}

For $k \leq 8$, the designs provided by ZM minimize the G- and G\textsubscript{2}-aberration criteria. When $k=9$, the minimum G-aberration design cannot be created by adding a row to an orthogonal array and differs from the minimum G\textsubscript{2}-aberration design presented by ZM. When $k$ is 10, 11 or 12, the minimum G- and G\textsubscript{2}-aberration designs differ, and the minimum G-aberration design cannot be obtained from an orthogonal array. The minimum G\textsubscript{2}-aberration designs we provide for these cases are better than those of ZM in terms of the G-aberration criterion. Finally, for $13 \leq k \leq 15$, we found designs that minimize both the G-aberration criterion and the G\textsubscript{2}-aberration criterion. The designs reported by ZM for these $k$ values are only optimal with respect to the G\textsubscript{2}-aberration criterion.

\spacingset{1}
\subsection{Minimally aliased DA designs with 18 runs} \label{ssec:bd_2mod4}
\spacingset{1.9}

Our enumeration and characterization of DA designs for run sizes that are two more than a multiple of four allowed us to identify many new 18-run designs that cannot be constructed by adding two rows to an orthogonal array. Of particular interest are the 18-run designs with more than 15 factors. To construct such designs with the method of \citet{galil1980d}, we would require 16-run orthogonal arrays with more than 15 factors, but such arrays do not exist. Therefore, HM do not provide benchmark designs for these cases.

In Table~\ref{tab:bd_2mod4_18}, we characterize the minimum G-aberration designs, the minimum G\textsubscript{2}-aberration designs and the designs presented by HM for $N_2$ equal to 18. For even $k$ values, we present the minimum G- and G\textsubscript{2}-aberration designs for both optimal forms of the information matrix. Otherwise, the layout of Table~\ref{tab:bd_2mod4_18} is the same as that of Table \ref{tab:bd_1mod4}. We limit ourselves to cases where $18 < 2^k$, because we believe that the designs for $18 > 2^k$ are less interesting from a practical point of view.

Cases for which HM provided a benchmark design are indicated by an `x' in the fifth column in Table \ref{tab:bd_2mod4_18}. HM do not discuss DA designs for even numbers of factors with an information matrix of the form $\Gamma_{(k/2,k/2+1)}$, even though these DA designs have a better F$_1$ vector.

\spacingset{1}
\begin{table}[!t]
%\footnotesize
  \centering
  \caption{Properties of the $k$-factor designs with minimum G-aberration (mG) and minimum G\textsubscript{2}-aberration (mG\textsubscript{2}) and of those presented by HM for run size $N_2$ equal to 18. nOA: design cannot be derived from an OA; J$_3^m$: maximum J$_3$ value; J$_4^m$: maximum J$_4$ value; $f$: frequency.}
    \begin{tabular}{ccccccccrr}
    \toprule
 $k$     & optimal form & mG & mG\textsubscript{2} & HM & nOA & (J$_3^m$, $f$) &  (J$_4^m$, $f$) & $C_2$    & $C_3$ \\
        \midrule
5     & $\Gamma_{3,3}$     & x     & x     & x     &       & (\phantom{0}2, \phantom{0}6)     & (\phantom{0}2, \phantom{00}3)     & 0.215 & 0.198 \\
6     & $\Gamma_{3,4}$     & x     &       &       &       & (\phantom{0}2, 12)    & (16, \phantom{00}2)     & 0.338 & 10.129 \\
6     &   $\Gamma_{3,4}$  &       & x     &       &       & (\phantom{0}2, 12)    & (18, \phantom{00}1)     & 0.338 & 10.053 \\
6     & $\Gamma_{4,3}$     & x     & x     & x     &       & (\phantom{0}2, 10)    & (16, \phantom{00}2)     & 0.348 & 10.125 \\
7     & $\Gamma_{4,4}$     & x     &       &       &       & (\phantom{0}2, 19)    & (18, \phantom{00}1)     & 0.521 & 24.396 \\
7     &  $\Gamma_{4,4}$   &       & x     & x     &       & (\phantom{0}2, 19)    & (18, \phantom{00}3)     & 0.521 & 24.368 \\
8     & $\Gamma_{4,5}$     & x     & x     &       &       & (\phantom{0}2, 31)    & (18, \phantom{00}3)     & 0.715 & 49.474 \\
8     & $\Gamma_{5,4}$     & x     & x     &       &       & (\phantom{0}2, 28)    & (18, \phantom{00}2)     & 0.728 & 49.036 \\
8     &  $\Gamma_{5,4}$    &       & x     & x     &       & (\phantom{0}2, 28)    & (18, \phantom{00}6)     & 0.728 & 49.036 \\
9     & $\Gamma_{5,5}$     & x     & x     &       & x     & (\phantom{0}8, \phantom{0}8)     & (18, \phantom{00}6)     & 9.751 & 56.379 \\
9     &  $\Gamma_{5,5}$    &       &       & x     &       & (14, \phantom{0}4)     & (18, \phantom{00}6)     & 10.272 & 50.580 \\
10    & $\Gamma_{5,6}$     & x     &       &       & x     & (\phantom{0}8, 12)    & (18, \phantom{00}4)     & 20.770 & 67.432 \\
10   &  $\Gamma_{5,6}$     &       & x     &       & x     & (10, \phantom{0}4)     & (18, \phantom{00}4)     & 18.674 & 77.517 \\
 10  & $\Gamma_{6,5}$     & x     &       &       &       & (\phantom{0}8, 16)    & (18, \phantom{0}10)    & 20.112 & 73.935 \\
10   & $\Gamma_{6,5}$     &       & x     &       & x     & (10, \phantom{0}2)     & (18, \phantom{00}6)     & 19.356 & 72.859 \\
10   & $\Gamma_{6,5}$    &       &       & x     &       & (14, \phantom{0}8)     & (18, \phantom{0}18)    & 20.112 & 73.935 \\
11    & $\Gamma_{6,6}$     & x     & x     &       &       & (\phantom{0}8, 24)    & (18, \phantom{00}8)     & 30.332 & 106.597 \\
 11  & $\Gamma_{6,6}$     &       & x     & x     &       & (14, 12)    & (18, \phantom{0}26)    & 30.332 & 106.597 \\
12    & $\Gamma_{6,7}$     & x     &       &       &       & (\phantom{0}8, 34)    & (18, \phantom{0}10)    & 43.287 & 155.131 \\
12   &  $\Gamma_{6,7}$    &       & x     &       &       & (10, \phantom{0}4)     & (18, \phantom{00}8)     & 41.779 & 154.681 \\
 12  & $\Gamma_{7,6}$     & x     & x     &       &       & (\phantom{0}8, 32)    & (18, \phantom{0}15)    & 40.769 & 159.113 \\
 12  & $\Gamma_{7,6}$     &       & x     & x     &       & (14, 16)    & (18, \phantom{0}39)    & 40.769 & 159.113 \\
13    & $\Gamma_{7,7}$     & x     & x     &       &       & (\phantom{0}8, 44)    & (18, \phantom{0}15)    & 56.427 & 223.618 \\
 13  &  $\Gamma_{7,7}$   &       & x     & x     &       & (14, 22)    & (18, \phantom{0}55)    & 56.427 & 223.618 \\
14    & $\Gamma_{7,8}$     & x     &       &       & x     & (10, \phantom{0}4)     & (18, \phantom{00}4)     & 76.865 & 287.218 \\
14   & $\Gamma_{7,8}$    &       & x     &       &       & (10, 12)    & (18, \phantom{0}21)    & 72.427 & 311.733 \\
14   & $\Gamma_{8,7}$     & x     & x     &       &       & (\phantom{0}8, 56)    & (18, \phantom{0}21)    & 72.338 & 311.934 \\
14   & $\Gamma_{8,7}$     &       & x     & x     &       & (14, 28)    & (18, \phantom{0}77)    & 72.338 & 311.934 \\
15    & $\Gamma_{8,8}$     & x     &       &       & x     & (10, \phantom{0}7)     & (18, \phantom{0}21)    & 93.345 & 408.413 \\
15   & $\Gamma_{8,8}$     &       & x     &       &       & (14, \phantom{0}7)     & (18, \phantom{0}21)    & 91.109 & 423.828 \\
15   &$\Gamma_{8,8}$      &       & x     & x     &       & (14, 35)    & (18, 105)   & 91.109 & 423.828 \\
16    & $\Gamma_{8,9}$     & x     & x     &       & x     & (14, \phantom{0}4)     & (18, \phantom{00}9)     & 118.142 & 524.532 \\
 16  & $\Gamma_{9,8}$     & x     & x     &       & x     & (10, 12)    & (18, \phantom{00}7)     & 118.463 & 520.085 \\
17    & $\Gamma_{9,9}$     & x  &   x     &       & x     & (18, \phantom{0}2)     & (18, \phantom{00}7)     & 143.997 & 680.003 \\
    \bottomrule
    \end{tabular}%
  \label{tab:bd_2mod4_18}%
\end{table}%
\spacingset{1.9}

When $k$ is 5 or 6, the HM design minimizes both the G- and the G\textsubscript{2}-aberration criterion. For $k=7$, our results confirm that the design of HM is a minimum G\textsubscript{2}-aberration design, but it differs from the minimum G-aberration design. In this case, the minimum G-aberration design can be constructed from an orthogonal array. For $k \in \{ 8,11,12,13,14\}$, the designs we report minimize both aberration criteria, while the designs provided by HM only minimize the G\textsubscript{2}-aberration criterion. For $k = 9$, the minimum G- and G\textsubscript{2}-aberration designs coincide and cannot be obtained from an orthogonal array. Consequently, the design provided by HM is suboptimal with respect to both aberration criteria. For $k=10$, the design of HM is also suboptimal with respect to both aberration criteria, but, in this case, the minimum G- and G\textsubscript{2}-aberration designs differ. The former can be obtained from an orthogonal array, while the latter cannot. When $k=15$, the minimum G- and G\textsubscript{2}-aberration designs differ, and the minimum G-aberration design cannot be obtained from an orthogonal array. The minimum G\textsubscript{2}-aberration designs we provide for these cases are better than those of HM in terms of the G-aberration criterion.

\spacingset{1}
\section{Conclusions} \label{sec:conclusion}
\spacingset{1.9}

DA designs for main-effects models have been studied thoroughly for numbers of runs that are multiples of four, but not for numbers of runs that are one or two more than a multiple of four. This is surprising, given the fact that the optimal forms of the information matrices of the DA designs are generally known for these kinds of numbers of runs. The only studies that paid detailed attention to numbers of runs that are one and two more than a multiple of four focused on DA designs that can be constructed by adding one or two rows to an orthogonal array. It turns out, however, that there are many DA designs that cannot be obtained by adding one or two rows to an orthogonal array, and that these DA designs sometimes perform better in terms of the commonly used G- and G$_2$-aberration criteria. As a result, there is added value in studying the complete set of DA designs, and not only those that can be constructed from orthogonal arrays.

We developed one algorithm to generate a complete set of non-isomorphic DA designs with one run more than a multiple of four and another, more complex, one to generate a complete set of non-isomorphic DA designs with two runs more than a multiple of four. The reason that the latter algorithm is more complex than the former is that there are two optimal forms for the information matrix of DA designs for even numbers of factors in the event the run size is two more than a multiple of four, and that DA designs for run sizes that are two more than a multiple of four possess two kinds of factor columns: columns whose elements sum to zero and columns whose elements sum to two. With the two algorithms, we obtained many non-isomorphic DA designs. A large portion of the designs we obtained cannot be constructed from orthogonal arrays.

We evaluated all the DA designs we obtained in terms of their aliasing properties. More specifically, we investigated the performance of our designs in terms of the G- and G$_2$-aberration criteria. On multiple occasions, we found designs in our newly obtained catalog that outperform the best known designs from the literature. As a result, we were able to advance the start of the art when it comes to designing experiments for main-effects estimation.

Our work implies that the use of heuristic algorithms for constructing D- or A-optimal designs for main-effects models is no longer necessary in the event the number of runs is smaller than 20 and one or two more than a multiple of four. There are two major weaknesses with the use of these heuristics. First, they cannot guarantee that a truly D- or A-optimal design will be produced. Second, in the event they do produce a D- or A-optimal design, it is unlikely that the resulting designs will minimize the aliasing between the main effects and the interaction effects and the aliasing among the interaction effects. Instead of resorting to heuristic algorithms, practitioners can now simply use the minimally aliased DA designs we identified in this paper.

For small numbers of runs, minimally aliased orthogonal arrays, which have numbers of runs that are multiples of four, have been studied in detail as well. {Therefore, in the event the number of runs is a multiple of four, heuristic algorithms for generating DA main-effects designs are not needed either.} As a result, the only type of run size that remains to be studied in detail is one that is three more than a multiple of four. This type of run size is more complex than the types studied in this paper, because multiple optimal forms exist for the information matrices of main-effects designs and because the optimal form of the information matrix {for D-optimal designs differs from the optimal form for A-optimal designs} \citep{gail1982construction, sathe1989optimal,sathe1990construction}.

% \section{BibTeX}

% We hope you've chosen to use BibTeX!\ If you have, please feel free to use the package natbib with any bibliography style you're comfortable with. The .bst file agsm has been included here for your convenience. 

\bibliographystyle{apalike}

\spacingset{1}
% \bibliography{short,bibliography.bib}
\bibliography{bibliography.bib}
\spacingset{1.9}

\end{document}